\documentclass[final,smallcondensed]{svjour3}
 
\usepackage{times}
\usepackage{graphics}
\usepackage{adjustbox}
\usepackage{blindtext}
\usepackage{pifont}
\usepackage{latexsym}
\usepackage{amsmath,amsfonts,amssymb,nicefrac}
\usepackage{booktabs,tabularx,multirow,caption}
\usepackage{graphicx,tikz}
\usepackage[noend]{algpseudocode}
\usepackage[ruled,vlined,linesnumbered,linesnumbered,lined,boxed,commentsnumbered]{algorithm2e}
\usepackage{todonotes}
\usepackage{skull}
\usepackage{rotating}
\usepackage{float}
\usepackage{subfigure,color,epsfig}
\usepackage{cite}
\usepackage{textcomp}
\usepackage{xcolor}
\def\BibTeX{{\rm B\kern-.05em{\sc i\kern-.025em b}\kern-.08em
    T\kern-.1667em\lower.7ex\hbox{E}\kern-.125emX}}

\newcommand{\p}{\ensuremath{\mathrm{\bf P}}}
\newcommand{\np}{\ensuremath{\mathrm{\bf NP}}}

\newcommand{\MM}{\ensuremath{\mathsf{MinMax}}}
\newcommand{\MS}{\ensuremath{\mathsf{MaxSum}}}
\newcommand{\mS}{\ensuremath{\mathsf{MinSum}}}

\def\cH{\mathcal H}

\usepackage{amsmath}
\usepackage{graphicx}

\begin{document}

%\title{The machine scheduling problems with conflict jobs: formulations and algorithms}

\title{Exact Algorithms for Scheduling Problems on Parallel Identical Machines with Conflict Jobs}

\titlerunning{Scheduling problems with conflict jobs}

\author{Minh Ho\`ang H\`a   \and Dinh Quy Ta \and Trung Thanh Nguyen*}

\authorrunning{H\`a et al.}

\institute{Minh Ho\`ang H\`a \and Trung Thanh Nguyen \at
              ORLab, Faculty of Computer Science, Phenikaa University, Hanoi 12116, Vietnam \\
             \email{thanh.nguyentrung@phenikaa-uni.edu.vn}           %  \\
           \and
           Dinh Quy Ta \at
           ORLab, Faculty of Information Technology, VNU University of Engineering and Technology, Hanoi, Vietnam
}

\date{Received: XXXX/XX/XX / Accepted: XXXX/XX/XX}

\maketitle

\begin{abstract}
Machine scheduling problems involving conflict jobs can be seen as a constrained version of the classical scheduling problem, in which some jobs are conflict in the sense that they cannot be proceeded simultaneously on different machines. This conflict constraint naturally arises in several practical applications and has recently received considerable attentions in the research community. In fact, the problem is typically $\np$-hard (even for approximation) and most of algorithmic results achieved so far have heavily relied on special structures of the underlying graph used to model the conflict-job relation. Our focus is on three objective functions: minimizing the makespan, minimizing the weighted summation of the jobs' completion time, and maximizing the total weights of completed jobs; the first two of which have been intensively studied in the literature. For each objective function considered, we present several mixed integer linear programming models and a constraint programming model, from which we can solve the problems to optimality using dedicated solvers. Binary search-based algorithms are also proposed to solve the makespan problem. The results of numerical experiments performed on randomly generated data sets with up to 32 jobs and 6 machines are reported and analysed to verify the performance of the proposed methods.
\end{abstract}

\keywords{Machine scheduling \and conflict job \and  mixed integer linear programming \and  constraint programming \and binary search.}

\section{Introduction}
\label{intro}
The classical scheduling problem on parallel machines (a.k.a  {\em minimum makespan}) deals with assigning a set of heterogenous jobs, each having a certain processing time, to a finite number of identical  machines so that the makespan is minimized, where the makespan is defined as the maximum load over all machines. While this problem has a long and rich history, its constrained versions, where additional requirements for the job assignment are attached, have also flourished in recent years. In this paper we investigate a salient and remarkable constrained variant   in which some jobs are conflict in the sense that they cannot be processed at the same time (on different machines).  For example, let's consider a scenario where there is a set of resources, each with a certain supply, and each job has a specified demand (or request) for each resource. A conflict occurs among jobs if their accumulative request of a resource does exceed its supply. The conflict constraint has also arisen in many other real-word applications such as the load balancing in a parallel computation \cite{BakerC96}, traffic intersection control \cite{Bell92,Bullock94}, VLSI and compiler design \cite{NicolosoSS99},  frequency assignment
in cellular networks and session management in local area networks  \cite{HalldorssonKPSST03}, resource-constrained scheduling \cite{GareyG75}. 

More recently, several new applications of the conflict constraints are introduced to demonstrate the situations where some jobs cannot be processed simultaneously to ensure the reliability of systems. In \cite{Ha2019}, the authors consider a scheduling problem motivated by a joint project with a mobile network operator, who needs to run various computer programs to  adjust configuration or  some parameters of the network. Each program affects a pre-determined group of  mobile  subscribers. Any two programs that affect two parts of the network which have common elements cannot be executed concurrently, and  this can be modeled by the conflict constraints.

Another application can be found in the ROADEF Challenge 2020 named ``Grid operation-based outage maintenance planning" proposed by Réseau de Transport d'Électricté (RTE), the electricity transmission system operator of France (see \cite{Ruiz}). To guarantee both electricity delivery and supply, RTE must ensure that the grid is correctly maintained. One of its difficult tasks is to schedule maintenance operations for its electricity lines. In reality, some lines are sometimes too close to each other to carry out corresponding interventions at the same time. This would significantly weaken the network, and would be dramatic in case another close line is disconnected during the interventions. Therefore, maintenance jobs corresponding to close lines cannot be performed simultaneously to ensure the safety of the whole power system. 

%MSCJ is of theoretical interest since it naturally generalizes the classical minimum makespan problem in combinatorial optimization. On the other hand, it also arises as important problems in a number of practical applications such as ... (see below for a detailed description of these applications). We classify the studied problems into ones with or without time constraint and the objectives in each class are different. 

A simple and widely-used tool for modeling the conflicts among jobs is to make use of an arbitrary {\em conflict graph}, whose vertices correspond to the jobs, and an edge between two vertices represents the conflict of these two jobs. It is worth noting that one can also consider the complement of a conflict graph, called {\em compatible} (or {\em agreement}) graph, in which there is an edge between two vertices if the corresponding jobs can be scheduled simultaneously. The presence of the constraint on conflict jobs makes a machine scheduling problem more general and harder to address, but worthy to study. The special case of edgeless graphs (i.e., graphs with no edges) captures exactly the classical setting where there are no conflict jobs. Beyond conflict graphs, the existing works in the literature have exhibited scenarios where each conflict may involve more than two jobs, and the conflict relationships can be expressed by a general structure such as hypergraphs  \cite{BlazewiczLK83}. On the other hand, there has been the case that the processing of some job must be done before the execution of another job can start, and such a conflict is unable to be represented in terms of (hyper-)graphs \cite{GareyG75}. Our focus in this work is on the scheduling jobs with pairwise conflicts. 

In the literature, machine scheduling problems involving conflict jobs have been studied under different objectives, including minimum makespan, minimum weighted-sum of completion times, and maximum total weights of completed jobs. The first objective, which aims to minimize the amount of time needed for processing all jobs, has been well-known for decades. The second objective has been an intuitively appealing objective and studied since the late 1980s and early 1990s, with a connection to the {\em sum coloring} problem \cite{Kubicka}. The problem of maximizing the third objective naturally arose as a generalized version of the classical {\em Knapsack problem} (KP) \cite{KelPfePis04}, where each job is associated with a profit and a weight (or length) that specifies the  amount of time that needed to  be processed. In this context, every two jobs are conflict and one may not have them all  executed on a single machine, assuming a deadline $T$ by which the chosen jobs must be finished. We therefore look for a subset of jobs of maximum total profits that can be scheduled on the machine. 

From a computational point of view, it is important to observe that the machine scheduling problem with each of the three objectives above is $\np$-hard, even for edgeless graphs, meaning that there are no polynomial time algorithms, unless $\p=\np$. For instance, the minimum makespan problem with two identical machines and an edgeless graph is at least as hard as the Partition problem \cite{gar-joh:b:int}. Also, the problem of maximizing  total weights of completed jobs on one machine with deadline $T$ is simply the classical Knapsack problem and is thus $\np$-hard. One can see that the conflict graph plays a major impact on the complexity of the corresponding machine scheduling problem. Most of the works so far have been therefore concentrated on specific structures of graphs for which one can either establish $\np$-harness results as well as approximation algorithms, or evenly provide polynomial time algorithms that solve the problem to optimality. An overview of such results is provided in  Section~\ref{sec:related-work}.

Despite various hardness and approximation results achieved for the machine scheduling problems with conflict jobs, there has not been much work focusing on developing exact algorithms for general conflict graphs. Although the drawback of such algorithms lies on its  exponential time, they may still be of theoretical interest as well as of a practical point of view (we refer to a survey \cite{Woeginger01} by Woeginger for more reasons of studying exact algorithms for $\np$-hard problems). In fact, for some realistic situations, really exact solutions are more preferred than  approximate ones. This motivates us to study in this work potential approaches for exactly solving  the machine scheduling problems with conflict jobs, which, along with known approximability behaviors, may help understanding better the hardness and the solvability of the problems.

\paragraph{Organization.} The remaining of the paper is structured as follows. In Section~\ref{subsec:basic} we formally introduce the problem models that are studied in this paper, followed by a brief overview of related works and our contribution. Section \ref{mip} introduces MILP formulations for the problems, while the CP formulations are described in Section \ref{cp}. The algorithms based on binary search for the minimum makespan problem are discussed in Section \ref{bs}. The performance comparison of the models is represented by experimental results in Section \ref{sec:exp}. Finally, Section \ref{sec:conclude} concludes our research.

%\textbf{HMH: should give examples, applications when each objective function arises (if possible)}.

%We propose modeling approaches and exact solution methods for a class of  machine scheduling problems with conflict jobs (MSCJ). ble.

%Moreover, it has been shown that it is even $\np$-hard to approximate the problem MSCJ to within a factor of $n^{\epsilon-1}$ for any $\epsilon>0$, where $n$ denotes the number of jobs.

\section{Problem model and related work}
\label{subsec:basic}

\subsection{Problem model}
Consider a set of $n$ jobs $J = \{1, \ldots, |J|\}$ and a set of parallel identical machines $M=\{1,\ldots,|M|\}$. Each job $i\in J$ has a processing time (or length) $p_i\in\mathbb{N}_{>0}$ and can be processed on at most one machine, without preemptions, that is, once a machine starts processing a job it continues this processing without interruptions until its completion. We call jobs of unit processing time unit-jobs. We denote by $p=(p_i)_{i\in J}$  the vector of job processing times. It is  assumed that each machine can process at most one job at a time. Some of jobs are conflict (or incompatible) in the sense that they cannot be scheduled simultaneously on different machines. The conflict relation between jobs is modeled in terms of an undirected graph $G = (J, E)$, on which each vertex corresponds to a job, and there is an edge connecting two vertices $i,i'$ if they represent two conflict jobs. In the following we will define an assignment (or schedule) of jobs to machines. It is assumed that time is discrete and every job and machine are available for processing at time zero.

\begin{definition}[Job assignment]
A job assignment (or schedule) is an assignment of time intervals on the machines
to the jobs in such a way that fulfills the following conditions: 
\begin{enumerate}
\item Each job $i$ is assigned an interval of length $p_i$ on one machine;
\item Every two intervals assigned to the same machine do not intersect;
\item For every two conflict jobs, their intervals (on different machines) do not intersect.
\end{enumerate}
For a job assignment $\pi$, we denote by $s_i$ the   starting point of an interval  on which a job $i$ is processed on a machine. The interval during which job $i$ is processed on a machine is $[s_i,s_i+p_i]$. Formally, the job assignment $\pi$ can be defined as a one-to-one mapping:
\begin{align*}
\pi: J& \rightarrow M \times [0,\mathcal{L}]\\
 i &\mapsto \pi(i)=(m_i,s_i),
\end{align*}
where $\mathcal{L}=\sum_{i\in J}p_i$, such that for every two jobs $i$ and $i'$ we have $[s_i,s_i+p_i]\cap [s_{i'},s_{i'}+p_{i'}]=\emptyset$ if they are conflict or if they are assigned to the same machine, i.e., $m_i=m_{i'}$. 
 %In case when $s_i=\infty$, we say that the job $i$ is not assigned to any machine.
  %In case when a job $j$ is not assigned to any machine due to time constraint we define $\pi_{ij}$ to be $\infty$.
\end{definition}

Our definition of job assignment is quite general as it covers both the cases with and without the deadline constraint. In the former case, where there is a deadline $T\le \mathcal{L}$, if there is an allocation $\pi(i)$ with $s_i\ge T$, it means that job $i$ was not assigned to any machine. We denote by $\Pi$ the set of all possible job assignments $\pi$.

In what follows, we formally define three machine scheduling problems that are studied in this paper.

\begin{definition}[$\MM$]
Given an instance problem $I=(J,M,p,G)$, compute an assignment $\pi$ such that jobs are completely  processed on machines and the maximum completion time over all machines is minimized, that is, 
\[
\min_{\pi\in\Pi}\left\{\max_{i\in J} \{s_i+p_i\}\right\},
\]
subject to the conflict constraint. 
In other words, we seek for a schedule in which the largest endpoint of an interval assigned to a job is as small as possible.
\end{definition}

\begin{definition}[$\mS$]
In this problem each job $i \in J$ is associated with a nonnegative weight (or cost) $w_i$ beyond the processing time $p_i$. Denote by $w=(w_i)_{i\in J}$ the vector of job weights.  Given an instance problem $I=(J,M,p,w,G)$, compute an assignment $\pi$ that minimizes the weighted-sum of completion times, i.e. 
\[
\min_{\pi\in\Pi}\left\{\sum_{i\in J} w_i\cdot (s_i+p_i)\right\},
\]
subject to the conflict constraint.  
%where $f_i (\pi)=s_i+p_i$ is the largest endpoint of the interval assigned to $i$ by the assignment $\pi$.
\end{definition}
One may assume the number of machines to be infinite, as considered in most of works in the literature. However, this exactly corresponds to a special case of our model defined here by setting the number of machines equal to the number of jobs.

\begin{definition}[$\MS$]
Similar to $\mS$, we assume that in this problem model there is a vector of job profits $r=(r_i)_{i\in J}$.  In addition, there is a deadline $T\le \mathcal{L}$, by which jobs need to be finished.  Given an instance problem  $I=(J,M,p,r,G,T)$, compute  an assignment $\pi$ that maximizes the total weights of completed jobs, that is,
 \[
\max_{\pi\in\Pi}\left\{\sum_{i\in J,s_i+p_i\le T} r_i\right\},
\]
subject to the conflict constraint. 
\end{definition}

%The considered $\mS$ problem has been motivated by a joint project with a mobile network operator. This operator needs to run various computer programs, which are referred to as Change Requests. The company has a team of technicians. Each of these technicians can execute at most one Change Request at a time on her/his computer without interruption. These Change Requests adjust configuration or some parameters of the network and each affects a pre-determined group of mobile subscribers. The Change Requests have different priorities (which can be modelled by associating with each Change Request a certain nonnegative weight). Any two  Change Requests that affect two parts of the network which have common elements cannot be executed concurrently. The programs must be run only during the working hours. So, it may be not possible to execute all the programs during one shift.

%\subsection{Application and Motivation}

\subsection{Relate work}
\label{sec:related-work}
This section aims to provide a review of the literature on the three scheduling problems with conflict jobs: $\MM$, $\mS$, and $\MS$.  Basically, adding the conflict constraint makes the scheduling problem much harder to solve, or even to approximate. Hence, most of work has paid attention on the special structures of the conflict graphs which can make the problem tractable or approximable.  To be more focused, the review below is separated for each considered problem. 

\paragraph{$\MM$ {\em problem}. } One can see that for the edgeless graph  (i.e. graph with no edges) all jobs are not conflict, and  $\MM$ comes down to the classical minimum makespan, which is $\np$-hard. Graham \cite{Gr66,Gr69} gave a greedy $3/4$-approximation algorithm for the problem. Hochbaum and Shmoys \cite{HochbaumS88} presented a polynomial time approximation scheme\footnote{For $\alpha\ge 1$, an $\alpha$-approximation algorithm for a minimization problem $P$ is a polynomial time algorithm that can produce a near-optimal solution whose value is at most $\alpha$ times the optimum. A PTAS is a $(1+\epsilon)$-approximation algorithm for any fixed constant $\epsilon\in(0,1)$. We refer to the textbook \cite{Vazi01} for more details of approximation algorithms.} (PTAS) and this is the best possible result for the problem, as it is strongly $\np$-hard (using a reduction from 3-Partition \cite{gar-joh:b:int}). Several exact methods have been proposed for solving the problem to optimality, including: cutting plane algorithm \cite{Mokotoff04},  branch-and-bound algorithms  \cite{DellAmicoM05,HaouariJ08},  branch-and-price algorithm \cite{DellAmicoIMM08}. 
%Hence, $\MM$ subsumes the classical minimum makespan problem as a special case and is thus $\np$-hard, that is, there is unlikely a polynomial time algorithm for exactly solving this problem, unless the standard assumption $\p=\np$ holds. Therefore, attempts have been made on the study of the complexity of the problem based on the special structures of the given conflict graph.  
For the case of general conflict graphs, $\MM$ with unit jobs was studied under the name {\em mutual exclusion scheduling} by Baker and Coffman \cite{BakerC96}. They proved that the problem is $\np$-hard in the strong sense if the number of machines is at least $3$, but is polynomially solvable for the two-machine case, by constructing a maximum matching for the complement of the conflict graph. The later positive result was extended by Even et al.  \cite{EvenHKR09} to the setting where the length of jobs is $1$ or $2$. Nevertheless, if one allows the job lengths to take values in domain $\{1,2,3\}$, then the problem becomes $\np$-hard \cite{BendraoucheB12}. Several other works have  considered unit jobs with different structures of the conflict graph \cite{alon83,Lonc91,BodlaenderJ93,Kaller95,Jansen03,EvenHKR09}. 
In fact, $\MM$ is $\np$-hard for permutation graphs \cite{Jansen03}, complements of comparability graphs \cite{Lonc91}, and for interval graphs \cite{BodlaenderJ93}. On the positive side, Baker and Coffman \cite{BakerC96} proved an interesting fact that $\MM$ with unit jobs is equivalent to the so-called $|M|$-{\em bounded coloring} problem\footnote{In a $\delta$-bounded coloring problem, one needs to find a  minimum coloring of a given graph such that every color class contains at most $\delta$ vertices.}, for which polynomial-time algorithms exist for trees (or forest)  \cite{BakerC96}, split graphs and complements of interval graphs \cite{Lonc91}, cographs and bipartite graphs \cite{BodlaenderJ93}, constant treewidth \cite{Kaller95}, and line graphs \cite{alon83}.

Approximability of $\MM$ has been investigated for special cases of job lengths. For example, the case with unit jobs can be expressed as the $|M|$-Set Cover problem, for which there exists an $(\cH_{|M|}-1/2)$-approximation algorithm \cite{DuhF97}, where $\cH_{|M|}=\ln(|M|)+\Theta(1)$, assuming that $|M|$ is constant. When $|M|$ is part of the input, one cannot approximate the problem to within a factor of $n^{1-\epsilon}$ \cite{Zuckerman06}, for any $\epsilon>0$, unless $\np\subset\text{DTIME}(n^{\log \log n})$. For the case when $p_i\in\{1,2,3\}$, there is  a $4/3$-approximation algorithm \cite{EvenHKR09} for $|M|=2$.

\paragraph{$\mS$ {\em problem}.}  As mentioned early, $\mS$ (in unweighted form with unit jobs) is related to the sum coloring problem (SCP) on graphs, which was first studied by Kubicka \cite{Kubicka} in her 1989 thesis (and independently by Supowit \cite{Supowit87} in the area of VLSI design). In this problem, colors are positive integers and the goal is to give to each vertex $i$ of the graph a color $p_i$  so that  adjacent vertices have different colors. In this context, the sum of colors given to the vertices is exactly the sum of completion times of the corresponding jobs. The minimum value of this sum is well known as the chromatic sum of the graph. A number of results have been proposed for SCP (see \cite{JinHH17} for a detailed survey). We briefly mention here  main algorithmic results. For general graphs, Bar-Noy et al. \cite{Bar-NoyHKSS00} presented an $\frac{|J|}{\log^2(|J|)}$-approximation algorithm, and proved that one cannot have an approximation factor better than $|J|^{1-\epsilon}$, for any $\epsilon>0$. In addition, they proved that the problem on line graphs is $\np$-hard, but can be approximated to a factor of $2$. $\np$-hardness (or inapproximability) results have been also established for planar graphs \cite{HalldorssonK02}, bipartite graphs and  perfect graphs \cite{Bar-NoyK98}, and interval graphs \cite{G01}. On the positive side, there is a PTAS for planar graphs \cite{HalldorssonK02} (and this even holds for non-unit jobs), while the approximation factors for bipartite graphs and interval graphs are, respectively, $\frac{27}{26}$ \cite{MJK04} and $1.796$ \cite{HalldorssonKS03}. When jobs are of arbitrary lengths, Gandhi et el. \cite{GandhiHKS08} gave improved factors of $7.682$ for line graphs, and of $11.273$ for interval graphs. Also, Bar-Noy et al. \cite{Bar-NoyK98} proposed a $2.8$-approximation factor for bipartite graphs and an $O(\log(|N|))$-approximation factor for perfect graphs. Other  $\np$-hardness results can be found in \cite{HoogeveenVV94,GiaroKMP01,Marx03}. Kubicka \cite{Kubicka} derived a polynomial time algorithm for trees, and this was then generalized to graphs of bounded treewidth  \cite{Jansen97}.  Other  conflict graph classes of proper intersection graphs of geometric objects (e.g. rectangles and disks) have been also studied in \cite{BorodinIYZ12}. 

To the best of our knowledge, exact methods for $\mS$ have been only proposed for its special case of SCP. The first compact Integer Programming (IP) model was given in \cite{SenDG92}, and its computational results together with another exponential-size IP model  were then studied in \cite{FuriniMMT18}. Very recently, Delle Donne et al. \cite{Donea20} have designed  the first branch-and-price algorithm to solve the SCP, using the column generation technique.

 %Hoogeveen et al. \cite{HoogeveenVV94} showed that $\mS$  on line graphs is weakly $\np$-hard. Marx \cite{Marx03}
%showed that $\mS$ of line graphs of trees is strongly $\np$-hard, even if all tasks
%have length $1$ or $2$. Further hardness results on restricted classes are given by Giaro
%et al. \cite{GiaroKMP01}. Approximation algorithms are developed for a class of fundamental graphs, including perfect graphs and bipartite graphs \cite{Bar-NoyHKSS00}, interval graphs and line graphs \cite{GandhiHKS08}, planar graphs \cite{HalldorssonK02}. The only known polynomial time algorithm is given for trees, which is obtained in \cite{HalldorssonKPSST03}. An interesting connection between $\mS$ and the classical minimum makespan is provided in \cite{GandhiHKS08}, where the authors prove that the former problem can be reduced to the latter one, via generalizing a technique of Queyranne and Sviridenko \cite{SVIRIDENKO02}. \textbf{HMH: we also note that several authors consider the case where the number of machines is infinite (cite here some papers). This version of the problem is related to the multi-coloring problem, which is well studied in the literature \cite{GandhiHKS08}. Our models proposed hereinafter can be easily adapted to solve this case by fixing the number of machines equal to that of jobs.}.

\paragraph{$\MS$ {\em problem}. } As a special case of $\MS$,  the Knapsack problem is known to be weakly $\np$-hard and admits a PTAS \cite{KelPfePis04}. For the case of unit jobs, the (weighted) $\MS$ with $T=1$ turns out to be equivalent to the (weighted) Maximum   Independent Set, which cannot be approximated to within a factor of $n^{1-\epsilon}$, for any $\epsilon>0$, \cite{H96}. 
H\`a et al.  \cite{Ha2019}  showed that the problem is polynomially solvable when there are two machines and the conflict graph is bipartite. In terms of exact methods, they formulated two mixed integer linear programming models to solve the problem in the general graph. We are not aware of any other work  done so far for $\MS$, except \cite{Ha2019}.

%As the best of our knowledge, there has not been much work devoted to the case of max-sum objective. In a very recent work \cite{Ha2019} the authors initiated the study of this problem and provided several algorithmic results as well as mixed integer based algorithms for the problem.

\paragraph{Other work.} We would like to mention that there is a number of articles in the literature concerning with scheduling problems of  {\em incompatible} jobs, in the sense that they cannot be performed on the {\em same machine} (see, for example \cite{BodlaenderJW94,KowalczykL17,MallekBB19}). This problem is totally different from our problems studied in this paper. 

\subsection{Our contribution}
%\paragraph{Our contribution. }

It is observed that the number of research works focusing on practically solving the problems with general graphs is quite limited. There are very few mathematical programming or constraint programming models, from which exact solution methods can be developed to solve the class of scheduling problems with conflict jobs. To the best of our knowledge, only in \cite{Ha2019}, two mixed integer linear programming models are proposed to solve the $\MS$ problem to optimality with MILP solvers. There are neither MILP nor CP models for the $\MM$ and $\mS$ problems. The lack of exact methods make the practically computational hardness of the problems be an open question that is interesting to investigate. To fill these gaps, we take the first step proposing and analysing in this research several mathematical models and constraint programming models. In particular, we introduce two new models, a mixed integer linear programming and a constraint programming, for the $\MS$ problem. We also adapt existing and new models for the remaining problems. Using dedicated model solvers, we develop exact algorithms that can solve the problems to optimality. A binary search technique is also used to reduce the size of the MILP models proposed for the $\MM$ problem and to develop a new exact method. The performance of all these exact methods is investigated on randomly generated problem instances and several usage recommendations are provided.

%\section{Scheduling problems with time constraint}

\section{Mixed integer linear programming models}
\label{mip}

%\subsection{Models}
%\label{sec:model}
\subsection{First formulation (F1)}
Let $[K] = \{1, 2, \ldots, K\}$ be the set of positions that a job can be processed on a machine. The cardinality of this set  can be determined as  $K=\lfloor \frac{T}{p_{\min}} \rfloor$, where $p_{\min} = \min_{i \in J} p_i$. The formulation denoted by (F1) and uses the following variables:
 \begin{itemize}
 \item  $x_{imk}$ : binary variables set to $1$ if job $i$ is done by machine $m$ at $k^{th}$ position
\item $s_i$: starting time of job $i$
\item  $z_{mk}$ : the completion time of a job at $k^{th}$ position by machine $m$
\item $y_l$: binary variables ($l = (i, j) \in C$) used to handle conflict constraint. It is set to $1$ if job $i$ is finished before starting time of job $j$, and set to $0$, otherwise.  
 \end{itemize}

The $\MS$ problem can be formulated as:
\begin{align}
    \text{(F1-$\MS$) \quad    Maximize} \; \; &\sum_{i \in J}\sum_{m \in M}\sum_{k \in K} r_ix_{imk} \label{f1-obj}\\
    \text{s.t.} \;\;\;  s_i + p_i \leq T + B( 1- &\sum_{m \in M}\sum_{ k \in K}x_{imk}) \quad \forall i \in J \label{f1-c1} \\
    \sum_{m \in M}\sum_{ k \in K}x_{imk}  \leq 1 &\quad \forall i \in J \label{f1-c2}\\
    \sum_{i\in J}x_{imk} \leq 1 &\quad \forall m \in M, k \in K \label{f1-c3} \\
    z_{m(k-1)} \leq z_{mk} &\quad \forall m \in M, k \in K, k > 1 \label{f1-c4}
\end{align}

\begin{align}
    s_i + p_i \leq z_{mk} + B( 1 - x_{imk}) &\quad \forall i \in J, m \in M, k \in K \label{f1-c5}\\
    s_i \geq z_{m(k-1)} - B( 1 - x_{imk}) &\quad \forall i \in J, m\in M, k\in K \label{f1-c6} \\
    s_i + p_i  \leq s_j + B y_{l} &\quad \forall l = (i, j) \in C \label{f1-c7}\\
    s_j + p_j \leq s_i + B(1-y_{l}) &\quad \forall l = (i, j) \in C \label{f1-c8}\\
    x_{imk} \in \{0, 1\} & \quad \forall i \in J, m \in M, k \in K \label{f1-c9}\\
    y_{l} \in \{0, 1\} & \quad \forall l = (i, j) \in C \label{f1-c10}\\
    s_i \in \mathbb{R^+} & \quad \forall i \in J \label{f1-c11}\\
    z_{mk} \in \mathbb{R^+} & \quad \forall m \in M, k \in K \label{f1-c12}
\end{align}
Objective function (\ref{f1-obj}) is to maximize the number of complete jobs with priorities as highest as possible. Constraints (\ref{f1-c1}) ensure the machine's capacity when they enforce every job must be finished before $T$. Constraints (\ref{f1-c2}) imply each job is processed on no more than one machine at only one position while constraints (\ref{f1-c3}) make sure that each machine can do at most once job at one position. Constraints (\ref{f1-c4}) indicate a job must be complete after its precedence. Constraints (\ref{f1-c5}) and (\ref{f1-c6}) express the relationship among $z$-variables and the others. Conflict constraints are satisfied by constraints (\ref{f1-c7}) and (\ref{f1-c8}). Finally, Constraints (\ref{f1-c9})-(\ref{f1-c12}) define variables' domain. In this formulation, $B$ is a very large number and can be estimated by the summation of the capacity and the execution time of all jobs which have a conflict with other jobs. The number of variables in (F1) is $|J|.|M|.|K| + |J| + |M|.|K|+ |C|$, and that of constraints is $2.|J|+ 2.|M|.|K| + |J|.|M|.|K| + 2.|C|$.

The $\mS$ problem can be modeled as follows:
\begin{align}
    \text{(F1-$\mS$)  \quad  \quad   Minimize} \; \; \sum_{i \in J} w_is_{i} \label{f1-p2-obj}\\
    \text{s.t.} \;\;\;  \sum_{m \in M}\sum_{ k \in K}x_{imk}  = 1 &\quad \forall i \in J \label{f1-p2-c2}\\
    (\ref{f1-c3}) - (\ref{f1-c12}) \nonumber
\end{align}

Objective function (\ref{f1-p2-obj}) is to minimize the summation of the weighted start processing time of all the jobs. This objective function is equivalent to the function minimizing the weighted completion time of all jobs. Indeed, the original objective function can be rewritten as $\sum_{i \in J}w_i(s_i + p_i) = \sum_{i \in J}w_is_i + \sum_{i \in J}w_ip_i$. Because the second term $\sum_{i \in J}w_ip_i$ is constant, we can remove it from the objective function without changing the optimal solution. Constraints (\ref{f1-p2-c2}) ensure that each job must be processed at one position on exactly one machine. Remaining constraints of the model are taken from formulation (F1-$\MS$).

To model the $\MM$ problem, we use an additional variable $T_{max}$ representing the completion time of the latest job. The problem can be stated as follows:
\begin{align}
    \text{(F1-$\MM$)  \quad \quad    Minimize} \; \; T_{max} \label{f1-p3-obj}\\
    \text{s.t.} \;\;\;  T_{max} \geq p_i + s_i \quad \forall i \in J \label{f1-p3-c2}\\
    (\ref{f1-p2-c2}), (\ref{f1-c3}) - (\ref{f1-c12}) \nonumber\\
    T_{max} \in \mathbb{R^+} \label{f1-p3-c3}
\end{align}
Objective function (\ref{f1-p3-obj}) is to minimize the completion time of the latest job. Constraints (\ref{f1-p3-c2}) imply the relationship of variable $T_{max}$ with other variables in the model. Constraint \ref{f1-p3-c3} defines the domain of variable $T_{max}$.
%%%%%%%% Chú ý: B có thể tính là min(T + tổng của p_i tất cả các job có conflict, tổng của p_i tất cả các job) sẽ có nhiều trường hợp nhỏ hơn em estimate

\subsection{Second formulation (F2)}
We now present the second MILP model that is denoted by (F2) for the problems. This model is proposed in \cite{Ha2019} and uses three types of variables as follows:
 
 \begin{itemize}
 \item $x_{im}$ : binary variables equal to 1 if job $i$ is done by machine $m$
 \item $s_{i}$: starting time of job $i$
 \item $y_{ij}$: binary variables equal to 1 if job $i$ is completed before starting time of job $j$
\end{itemize}

The problem can be stated as:
\begin{align}
\text{(F2-$\MS$) \quad  Maximize} \; \; &\sum_{i \in J}\sum_{m \in M} r_ix_{im} \label{f2-obj}\\
\text{s.t.} \;\;\; s_i + p_i \leq T + &B( 1 - \sum_{m \in M}x_{im})\quad \forall i \in J \label{f2-c1}\\
\sum_{m \in M}x_{im}  \leq 1 &\quad \forall i \in J \label{f2-c2}\\
s_{i} + p_{i} \leq s_{j} + B( 1 - & y_{ij}) \quad \forall i, j \in J, i\neq j \label{f2-c3}\\
y_{ij} + y_{ji} \geq x_{im} + x_{jm} - 1 &\quad \forall i, j \in J, i \neq j,  m \in M \label{f2-c4}\\
y_{ij} + y_{ji}  \geq 1 &\quad \forall (i, j) \in C \label{f2-c5}\\
x_{im} \in \{0, 1\} &\quad \forall i \in J, m \in M \label{f2-c6}\\
s_i \in \mathbb{R^+} &\quad \forall i \in J \label{f2-c7}\\
y_{ij} \in \{0, 1\} &\quad \forall i, j \in J \label{f2-c8}
\end{align}

Similarly to (F1), constraints (\ref{f2-c1}) imply capacity requirements. Constraints (\ref{f2-c2}) ensure that each job is processed on no more than one machine. Constraints (\ref{f2-c3}) express the relationship between variables $t$ and $y$. Constraints (\ref{f2-c4}) say that if two jobs $i$ and $j$ are processed on the same machine, one job must be completed before starting time of another. They are used to avoid the overlap among jobs processed on the same machine. Conflict constraints are assured by constraints (\ref{f2-c5}). Constraints (\ref{f2-c6})-(\ref{f2-c8}) define variables' domain. Similarly to (F1), $B$ in Constraints (\ref{f2-c2}) and (\ref{f2-c4}) is a very large number and can be estimated by the summation of the capacity and the execution time of all jobs which have a conflict with other jobs. The number of variables in the formulation (F2) is $|J|^2 + |M|.|J| + |J|$, while the number of constraints is $|J|^2 +|J|^2.|M| + |C| + 2.|J|$.

Using this formulation, the $\mS$ problem can be formulated as:
\begin{align}
\text{(F2-$\mS$) \quad  Minimize} \; \; &\sum_{i \in J} w_is_{i} \label{f2-p2-obj}\\
\text{s.t.} \;\;\; \sum_{m \in M}x_{im}  =  1 &\quad \forall i \in J \label{f2-p2-c2}\\
(\ref{f2-c3})-(\ref{f2-c8}) \nonumber
\end{align}

Objective function (\ref{f2-p2-obj}) is to minimize the weighted processing start time of all the jobs. Constraints (\ref{f2-p2-c2}) ensure that each job must be processed on exactly one machine. Remaining constraints of the model are taken from formulation (F2-$\MS$). The formulation proposed for the third problem is:
\begin{align}
\text{(F2-$\MM$) \quad  Minimize} \; \; T_{max} \label{f2-p3-obj}\\
\text{s.t.} \;\;\; (\ref{f1-p3-c2}), (\ref{f2-p2-c2}), (\ref{f2-c3})-(\ref{f2-c8}) \nonumber
\end{align}

\subsection{Third formulation (F3)}
We now introduce a new formulation for the problems. It uses three types of variables to model the first problem as follows:
\begin{itemize}
    \item $s_i$ : non-negative variable representing the start time of job $i$.
    \item $x_{imt}$: binary variable equal to one if and only if job $i$ is processed at time $t$ ($t \in \{1, ..., T\}$) on machine $m$.
    \item $z_i$: true if and only if job $i$ is scheduled.
\end{itemize}
  The $\MS$ problem can be formulated as:
    \begin{align}
        \text{(F3-$\MS$)    Maximize} \quad &\sum_{i \in J}r_iz_i \label{f3-obj}\\
            %moi may tai moi thoi diem chi thuc hien duoc toi da 1 job%
            \text{s.t.} \;\;\; \sum_{i \in J} x_{imt} \leq 1 &\quad \forall m \in M, \forall t \in \{1, ..., T\} \label{f3-c1}\\
            %moi job chi thuc hien toi da boi 1 may tai moi thoi diem%
            \sum_{m \in M} x_{imt} \leq 1 &\quad \forall i \in J, \forall t \in \{1, ..., T\} \label{f3-c2}\\
            %define xjmt%
            \sum_{m \in M} \sum_{t \in \{1, ..., T\}} x_{imt} = p_iz_i &\quad \forall i \in J\ \label{f3-c3}\\
            t - s_i \geq B(\sum_{m \in M}x_{imt} - 1) &\quad \forall t \in \{1, ..., T\}, \forall i \in J \label{f3-c4}\\
            s_i+p_i -t-1 \geq B(\sum_{m \in M}x_{imt} - 1) &\quad \forall t \in \{1, ..., T\}, \forall i \in J \label{f3-c5}\\
            %conflict ct%
            \sum_{m \in M} (x_{imt} + x_{jmt}) \leq 1 &\quad (i,j) \in C, t \in \{1, ..., T\} \label{f3-c6}\\
            x_{imt} \in \{0, 1\} &\quad \forall i \in J, m \in M, t \in \{1, ..., T\} \label{f3-c7}\\
            s_i \in \mathbb{R^+} &\quad \forall i \in J \label{f3-c8}\\
            z_{i} \in \{0, 1\} &\quad \forall i \in J \label{f3-c9}
    \end{align}

Objective function (\ref{f3-obj}) has the same meaning as functions (\ref{f1-obj}) and (\ref{f2-obj}). Constraints (\ref{f3-c1}) ensure that on each machine and at a time there is at most one job is processed. Constraints (\ref{f3-c2}) imply that at a time, each job is processed on at most one machine. Constraints (\ref{f3-c3}) state that, if job $i$ is scheduled, its total processing time must be equal to $p_i$. The two next constraints have the meaning as follows: if a job is processed at time $t$, its start processing time must be less than or equal to $t$ (Constraints \ref{f3-c4}) and its completion time must be larger or equal to $t$ (Constraints \ref{f3-c5}). Constraints (\ref{f3-c6}) indicate the conflict requirements. The remaining constraints define the variables' domains. 

%The number of variables of this formulation is $|J|.|M|.|T| + 2.|J|$ and that of constraints is $|M|.|T|+3.|J||T|+|C||T| + |J|$
In the two remaining problems, parameter $T$ is not defined. However, it is necessary for the definition of variables $x_{imt}$. Here, we estimate it by a trivial upper bound on the completion time of the latest job $T = \sum_{i \in J}p_i$.

The $\mS$ problem is formulated as:
    \begin{align}
        \text{(F3-$\mS$) \quad \quad  Minimize} \; \; &\sum_{i \in J} w_is_{i} \label{f3-p2-obj}\\
            %moi may tai moi thoi diem chi thuc hien duoc toi da 1 job%
            %\sum_{i \in J} x_{imt} \leq 1 &\quad \forall m \in M, \forall t \in T \label{f3-p2-c1}\\
            %moi job chi thuc hien toi da boi 1 may tai moi thoi diem%
            %\sum_{m \in M} x_{imt} \leq 1 &\quad \forall i \in J, \forall t \in T \label{f3-p2-c2}\\
                     % 2 constraints bi sai a, thay cai nay vao%
           \text{s.t.} \;\;\; \sum_{m \in M} \sum_{t \in \{1, ..., T\}} x_{imt} &= p_i \quad \forall i \in J \label{f3-p2-c1}\\
             (\ref{f3-c1}), (\ref{f3-c2}), & (\ref{f3-c4})-(\ref{f3-c9}) \nonumber
    \end{align}
Constraints (\ref{f3-p2-c1}) state that the total processing time of job $i$ is must be equal to $p_i$. Finally, we can easily model the $\MM$ problem as follows:
    \begin{align}
        \text{(F3-$\MM$) \quad \quad  Minimize} \; \; T_{max} \label{f3-p2-obj}\\
            %moi may tai moi thoi diem chi thuc hien duoc toi da 1 job%
           \text{s.t.} \;\;\; (\ref{f1-p3-c2}), (\ref{f3-c1}), (\ref{f3-c2}), (\ref{f3-c4})-(\ref{f3-c9}) \nonumber
    \end{align}
    
\subsection{Comparison of three MILP models}   

In general, the performance of a MILP formulation partially depends on its size, which is represented by the number of constraints and variables. In this section, we compare the three MILP formulations in terms of the number of constraints and variables. The size of the MILP models depending on the input data is presented in Table \ref{ComparModel}.

\begin{table}[h]
\begin{tabular}{|c|c|c|c|}
\hline
                  & (F1)                                       & (F2)                               & (F3)                              \\ \hline
\# of Variables   & $\mathcal{O}(|J|.|M|.|K| + |C|)$       & $\mathcal{O}(|J|^2 + |M|.|J|)$          & $\mathcal{O}(|J|.|M|.T)$           \\ \hline
\# of Constraints & $\mathcal{O}(|J|.|M|.|K| + |C|)$ & $\mathcal{O}(|J|^2.|M| + |C|)$ & $\mathcal{O}(|M|.T+|J|T+|C|T)$ \\ \hline
\end{tabular}
\caption{Comparison of three MILP models on their size}
\label{ComparModel}
\end{table}

As can be seen in the table above, the number of constraints and variables in all three formulations depends on the numbers of jobs $|J|$ and machines $|M|$ of the instance. However, the size of MILP model based on formulation (F2) depends more on the number of jobs because the numbers of constraints and variables in (F2) quadratically increase with $|J|$. The number of conflict job pairs $|C|$ has a significant impact on the number of constraints of all three formulations, but the impact is larger on the first formulation when the number of variables in (F1) also depends on $|C|$. The value of $T$, which is used to compute $|K|$, affects the size of formulations (F1) and (F3). 

\section{Constraint programming models}
\label{cp}
In this section, we present constraint programming models for the problems. First, it is worth mentioning that unlike MILP formulations, there is no standard in CP formulation because it strongly depends on each CP package. In this study, we formulate the models using generic keywords and syntaxes of IBM ILOG CP Optimizer \cite{CPO}. Our model uses the following variables:
\begin{itemize}
    \item $Itv_i$: interval variable that represents the time interval of size $p_i$ job $i \in J$ is proceeded.
    \item $ItvAlt_{im}$: optional interval variable that represents the time interval machine $m \in M$ proceeds job $i \in J$.
    \item $Seq_m$: sequence variable that represents all working time intervals of machine $m \in M$. 
\end{itemize}

The $\MS$ problem can be stated as:
\begin{align}
    \text{(CP-$\MS$) \quad   Maximize} &\sum_{i \in J}r_i \texttt{presenceOf}(Itv_i) \label{cp-obj}\\
            %moi job chi thuc hien boi mot may%
    \text{s.t.} \;\;\; \texttt{alternative}(Itv_i, ItvAlt_{im}: m \in M) &\quad \forall i \in J \label{cp-c1}\\
            %cac job phai hoan thanh truoc capacity%
    \texttt{endOf}(Itv_i) &\leq T \quad \forall i \in J \label{cp-c2}\\
            %cac job trong mot may khong overlap%
    \texttt{noOverlap}(Seq_m) & \quad \forall m \in M \label{cp-c3}\\
            %conflict%
    \big(\texttt{endOf}(Itv_i) \leq \texttt{startOf}(Itv_j) \big) \; \texttt{OR} \; \big( \texttt{endOf}(Itv_j) &\leq \texttt{startOf}(Itv_i) \big) \; (i,j) \in C \label{cp-c4}
    %\texttt{noOverlap}([Itv_i, Itv_j]) & \quad (i,j) \in C \label{cp-c4}\\
    %\text{interval}\; Itv_i, \; \text{optional} \quad &\forall i \in J \label{cp-c5}\\
    %\text{interval}\; ItvAlt_{im}, \; \text{optional}, \;\text{size} = p_i \quad &\forall i \in J, m \in M \label{cp-c6}\\
    %\text{sequence} \; Seq_m \quad &\forall m \in M \label{cp-c7}
\end{align}

The objective function is represented by (\ref{cp-obj}). Constraints \ref{cp-c2}) are the capacity constraints. Next, constraints (\ref{cp-c3}) make sure that jobs processed on each machine are not overlapped while the conflict constraints are ensured by constraints (\ref{cp-c4}). Here, the no-overlap constraint \texttt{noOverlap}($s$) on a sequence variable $s$ states that permutation $s$ defines a chain of non-overlapping intervals, any interval in the chain being constrained to end before the start of the next interval in the permutation.

In $\mS$ and $\MM$ problems, all the jobs must be processed, the variable $Itv_i$ ($i \in J$) is therefore not optional. The problems are formulated as follows:
\begin{align}
    \text{(CP-$\mS$) \quad \quad Minimize} \; \; &\sum_{i \in J} w_i \texttt{startOf}(Itv_i) \\
     \text{s.t.} \;\;\; &(\ref{cp-c1}),(\ref{cp-c3}),(\ref{cp-c4}) \nonumber
     %\text{interval} & \;\; Itv_i \quad \forall i \in J \label{cp-p2-c1} 
\end{align}
\begin{align}
    \text{(CP-$\MM$) \quad \quad  Minimize} \; \; &\texttt{max}_{i \in J} \texttt{endOf}(Itv_i)\\
     \text{s.t.} \;\;\; &(\ref{cp-c1}),(\ref{cp-c3}),(\ref{cp-c4}) \nonumber
\end{align}

\section{Binary search based algorithms for $\MM$ problem}
\label{bs}
As analysed above, the number of variables and constraints in formulations (F1) and (F3) heavily depends on the completion time of the latest job $T$. For the $\MM$ problem with the min-max objective function, we can perform the binary search to reduce the size of $T$ as presented in Algorithm \ref{algo1}. Let $T_{UB}$ and $T_{LB}$ denote an upper bound and a lower bound of $T$. Recall that the upper bound for any instance can be computed as $T_{UB} = \sum_{i \in J}p_i$. As a result, the lower bound $T_{LB}$ can be estimated trivially as $T_{LB} = \lceil  \frac{T_{UB}}{m} \rceil$. These bounds provide an initial interval for the binary search. The algorithm then iteratively solves model (F1-$\MM$) or (F3-$\MM$) with $T = T_{MID}$ representing the middle of the current search interval. If a feasible solution $S$ with the gap returned by MILP solver less than a given threshold $\gamma$ is found, the upper bound is updated to $obj(S) -1$, since now it is known that a solution with objective value $obj(S)$ can exist, but it is not known if a solution with less completion time exists yet. If no feasible solution is found, the lower bound is set to $T_{MID} + 1$, as there cannot exist any solution with the completion time less than $T_{MID} + 1$. After the loop, the algorithms simply returns the last solution found. This is sufficient, as the final iteration of the algorithm already proved that any solution with one less unit of the completion time cannot exist.

It is worth mentioning that the value of $\gamma$ has a significant impact on the convergence speed of the algorithm. If $\gamma$ is set to a relatively large number, the running time of an iteration would be quick, but the number of iterations could be very large. On the other hand, if $\gamma$ is set to a small value, e.g,. $\gamma$ is set to 0 corresponding to the case where the MILP models mentioned in Line 5 of Algorithm \ref{algo1} are required to be solved to optimality, we can stop the search process as soon as the problem at an iteration is successfully solved. In this case, the required number of solved MILP models is the least, but the running time required to process each iteration could be very large. Therefore, the value of $\gamma$ needs to be carefully selected so that we get a good trade-off between the number of required iterations and the running time spent at each iteration. By experiment, $\gamma$ is set to 10\% in this study.

\begin{algorithm}[H]
\label{algo1}
\SetAlgoLined
    $T_{UB} \gets \sum_{i\in J}p_i;$\\
    $T_{LB} \gets \lceil \frac{T_{UB}}{m} \rceil;$\\
    \While{$T_{LB} \leq T_{UB}$} {
        $T_{MID} = \lfloor(T_{UB} + T_{LB}) / 2\rfloor;$\\
        Solve the corresponding formulation (F1-$\MM$) with $K=\lfloor \frac{T_{MID}}{p_{\min}} \rfloor$ or (F3-$\MM$) with $T = T_{MID}$;\\
        \If{a feasible solution S is found}{ 
        %with the gap less than or equal to $\gamma$  is found}{
            %$Sol \gets$ Save the solution;\\
            $T_{UB} \gets obj(S) - 1$;\\
        }
        \Else{
            $T_{LB} \gets T_{MID} + 1$;\\
        }
    }
    \Return $S$;\\
    \caption{Binary Search Algorithm}
\end{algorithm}

%\begin{algorithm}[H]
%\SetAlgoLined
   % $UB \gets sum( priority );$\\
   % $LB \gets UB / m;$\\
   % \While{$LB \leq UB$} {
   %     $MID = (UB + LB) / 2;$\\
   %     $temp = check(MID, time\_limit - consumed\_time, gap);$\\
    %    \uIf{$temp.status == Solution\_Found$}{
    %        $ANS \gets temp.value;$\\
     %       $UB \gets temp.value - 1$;\\
      %  }
      %  \Else{
       %     \If{$!temp.is\_time\_limit\_exceed$} {
        %        $LB \gets MID + 1$;\\
       %     }
       % }
   % }
%,    \Return $ANS$;\\
   % \caption{Optimal Binary Search Algorithm}
%\end{algorithm}

\section{Experimental Results}
\label{sec:exp}
In this section, we discuss all our computational experiments with the models and solution methods proposed above. All algorithms described in this paper are implemented in C++. The MILP models are solved using CPLEX 12.10.0 and the CP model is solved with CP Optimizer 12.10.0. All the experiments are done on a CPU with AMD Ryzen 7 3700X 3.6 Ghz, running on Windows 10. To make ease for the result reproduction, we use a single thread to solve the models. For a fair comparison, all the methods are managed to run in a time limit of 1 hour on each instance.

A total of 432 instances are created using a procedure partly similar to the one proposed in \cite{Ha2019}. They are randomly generated with following parameters:
\begin{itemize}
    \item The number of jobs $|J|$ is set to \{16, 24, 32\}
    \item The number of machines $|M|$ is set to \{2, 4, 6\}
    \item For the $\MS$ problem, the capacity  $T$ is set to \{10, 20, 40, 80\}.
    \item The priority of each job $i$ is set to a random integer from 1 to 5.
    \item The weight of each job $i$ is set to a random integer from 1 to 5.
    \item Two jobs are consecutively considered as conflict with each other in a random fashion until the number of conflict pairs is equal to $\alpha.n$ where $\alpha$ is the conflict rate and set to one of four values $\{1, 2, 4, 8\}$. This is generated such that conflict jobs of larger $\alpha$ include those of smaller $\alpha$.
    \item The processing time of each job is a random integer from $ \lceil T/4 \rceil $ to $ \lfloor T/2 \rfloor $ (setting 1), from 1 to $\lfloor T/2 \rfloor $ (setting 2), and from 1 to $\lfloor T/8 \rfloor $ (setting 3), where $T$ in all three problems is set to \{10, 20, 40, 80\}. We note that, for the $\mS$ and $\MM$ problems, we use the machine capacity $T$ from the $\MS$ to create the processing time of each job. 
\end{itemize}

To investigate the impact of the instances' characteristics on the performance of the models, we report the average values of comparison criteria for each group of instances depending on the number of jobs, the number of machines, the machine capacity, the conflict rate, and the processing time settings. Three main criteria used for the comparison are: the running time in seconds (Time), the gap between the best lower and upper bounds (Gap), and the number of instances solved to optimality (Opt).

\subsection{Results for the $\MS$ problem}

We compare four formulations: (F1-$\MS$), (F2-$\MS$), (F3-$\MS$), and (CP-$\MS$) on the instances of $\MS$ problem. The results are presented in Table \ref{tab-p1}. We can derive several observations as follows:
\begin{itemize}
    \item In overall, CP performs the best when it can solve to optimality 397 over 432 instances, followed by (F3), (F1), and (F2) with 370, 237, and 198 successfully solved instances, respectively. We also note that there are three cases where (F3) can solve more instances to optimality than (CP): the instances with 16 jobs and those with machine capacity equal to 10 or 20.
    \item The performance of all the formulations tends to decrease when the number of jobs increases. Formulation (F2) is the most affected. When the number of jobs increases to 32, the number of successfully solved instances of (F2) is the least. This could be explained by the fact that the number of variables and constraints of (F2) is a quadratic function of the number of jobs. 
    \item The performance of (F3) and (CP) is better when the number of machines is small. However, formulation (F2) tends to work better when the number of machines increases. The impact of the number of machines on the performance of (F1) is not really clear.
    \item The machine capacity has less impact on the performance of formulations (F1), (F2), and (CP). But the high values of the capacity lead to significant reduction on the number of successfully solved instances for formulation (F3). This phenomenon is predictable because the number of variables and constraints of (F3) depends on the machine capacity.
    \item Although the number of constraints of all formulations increases when the conflict rate increases, the performance of the four formulations is insensitive with the conflict rate. Except for formulation (F1), when the conflict rate is set to the highest value, the number of successfully solved instances drastically reduces. This could be due to the fact that only in formulation (F1), the number of variables increases when the conflict rate increases.
    \item Setting 3 of the processing time tends to be easier to solve than the others for the MILP formulations. On tested instances, the performance of CP is insensitive to the processing time settings.
\end{itemize}

\begin{table}[!htbp]
\resizebox{\columnwidth}{!}{
\begin{tabular}{|c|c|c|c|c|c|c|c|c|c|c|c|c|}
\hline
\multirow{2}{*}{Instance} & \multicolumn{3}{c|}{F1} & \multicolumn{3}{c|}{F2} & \multicolumn{3}{c|}{F3} & \multicolumn{3}{c|}{CP} \\ \cline{2-13} 
                          & Time    & Gap     & Opt & Time    & Gap     & Opt & Time    & Gap     & Opt & Time     & Gap    & Opt \\ \hline
\multicolumn{13}{|c|}{Number of jobs}                                                                                             \\ \hline
16                        & 1592.5  & 0.0942  & 85  & 495.3   & 0.0095  & 130 & 290.0   & 0.0019  & 138 & 275.816  & 0.0076 & 133 \\ \hline
24                        & 2111.2  & 0.1498  & 67  & 1973.6  & 0.1939  & 70  & 799.3   & 0.0188  & 119 & 197.816  & 0.0009 & 138 \\ \hline
32                        & 2482.1  & 0.2526  & 46  & 2694.7  & 0.4690  & 37  & 1004.9  & 0.0465  & 113 & 500.618  & 0.0043 & 126 \\ \hline
\multicolumn{13}{|c|}{Number of machines}                                                                                         \\ \hline
2                         & 2194.4  & 0.2327  & 64  & 2198.8  & 0.4708  & 59  & 315.2   & 0.0014  & 141 & 75.931   & 0.0024 & 141 \\ \hline
4                         & 2236.7  & 0.1536  & 58  & 1694.4  & 0.1419  & 80  & 799.4   & 0.0255  & 118 & 366.145  & 0.0042 & 131 \\ \hline
6                         & 1754.7  & 0.1103  & 76  & 1270.4  & 0.0597  & 98  & 979.6   & 0.0402  & 111 & 532.173  & 0.0062 & 125 \\ \hline
\multicolumn{13}{|c|}{Machine capacity}                                                                                           \\ \hline
10                        & 2288.6  & 0.2084  & 42  & 1943.0  & 0.2353  & 53  & 51.5    & 0.0000  & 108 & 191.127  & 0.0062 & 103 \\ \hline
20                        & 2123.2  & 0.1408  & 49  & 1722.5  & 0.2166  & 59  & 262.6   & 0.0025  & 101 & 415.218  & 0.0023 & 97  \\ \hline
40                        & 1954.1  & 0.1691  & 53  & 1618.4  & 0.2313  & 63  & 823.1   & 0.0212  & 90  & 359.349  & 0.0051 & 98  \\ \hline
80                        & 1881.8  & 0.1438  & 54  & 1601.0  & 0.2134  & 62  & 1655.0  & 0.0657  & 71  & 333.306  & 0.0035 & 99  \\ \hline
\multicolumn{13}{|c|}{Conflict rate}                                                                                              \\ \hline
1                         & 1920.7  & 0.1454  & 52  & 1860.0  & 0.2835  & 55  & 577.2   & 0.0066  & 95  & 287.626  & 0.0016 & 100 \\ \hline
2                         & 1800.8  & 0.1388  & 57  & 1622.0  & 0.2497  & 61  & 555.7   & 0.0067  & 96  & 409.446  & 0.0030 & 97  \\ \hline
4                         & 1754.9  & 0.1242  & 61  & 1422.1  & 0.1898  & 66  & 734.7   & 0.0137  & 92  & 233.812  & 0.0023 & 103 \\ \hline
8                         & 2771.4  & 0.2537  & 28  & 1980.7  & 0.1735  & 55  & 924.6   & 0.0625  & 87  & 368.116  & 0.0101 & 97  \\ \hline
\multicolumn{13}{|c|}{Processing time settings}                                                                                   \\ \hline
1                         & 2589.3  & 0.2075  & 48  & 2306.3  & 0.4699  & 57  & 911.1   & 0.0310  & 114 & 531.768  & 0.0046 & 126 \\ \hline
2                         & 3071.5  & 0.2782  & 25  & 2181.9  & 0.1834  & 63  & 1057.8  & 0.0361  & 112 & 167.461  & 0.0006 & 138 \\ \hline
3                         & 525.0   & 0.0109  & 125 & 675.3   & 0.0192  & 117 & 125.2   & 0.0000  & 144 & 275.021  & 0.0076 & 133 \\ \hline
\multicolumn{13}{|c|}{}                                                                                                           \\ \hline
Total                     & 2061.9  & 0.1655  & 198 & 1721.2  & 0.2241  & 237 & 698.0   & 0.0224  & 370 & 324.750  & 0.0043 & 397 \\ \hline
\end{tabular}
}
\caption{Comparison of different formulations for the $\MS$ problem}
\label{tab-p1}
\end{table}

\subsection{Results for the $\mS$ problem}
%F3 co 6 instances ko ra%
\begin{table}[!htbp]
\resizebox{\columnwidth}{!}{
\begin{tabular}{|c|c|c|c|c|c|c|c|c|c|c|c|c|}
\hline
\multirow{2}{*}{Instance} & \multicolumn{3}{c|}{F1} & \multicolumn{3}{c|}{F2} & \multicolumn{3}{c|}{F3} & \multicolumn{3}{c|}{CP} \\ \cline{2-13} 
                          & Time    & Gap     & Opt & Time    & Gap     & Opt & Time    & Gap     & Opt & Time     & Gap    & Opt \\ \hline
\multicolumn{13}{|c|}{Number of jobs}                                                                                             \\ \hline
16                        & 2821.8  & 0.3322  & 32  & 2015.2  & 0.1953  & 68  & 2942.7  & 0.5881  & 28  & 1566.780 & 0.3686 & 87  \\ \hline
24                        & 3520.5  & 0.5246  & 4   & 3304.3  & 0.4507  & 14  & 3297.1  & 0.7779  & 13  & 2877.710 & 0.6860 & 36  \\ \hline
32                        & 3600.0  & 0.6824  & 0   & 3600.0  & 0.6427  & 0   & 3354.7  & 0.8130  & 11  & 3600.001 & 0.9591 & 0   \\ \hline
\multicolumn{13}{|c|}{Number of machines}                                                                                         \\ \hline
2                         & 3580.0  & 0.7094  & 1   & 3493.5  & 0.6320  & 6   & 3213.7  & 0.7323  & 17  & 3199.712 & 0.8320 & 20  \\ \hline
4                         & 3329.8  & 0.4902  & 11  & 3031.2  & 0.3984  & 26  & 3229.0  & 0.7347  & 15  & 2575.574 & 0.6370 & 46  \\ \hline
6                         & 3032.5  & 0.3397  & 24  & 2394.8  & 0.2583  & 50  & 3151.8  & 0.7082  & 20  & 2269.203 & 0.5447 & 57  \\ \hline
\multicolumn{13}{|c|}{Machine capacity}                                                                                           \\ \hline
10                        & 3241.6  & 0.5095  & 12  & 2963.1  & 0.4336  & 21  & 2452.7  & 0.4973  & 36  & 2715.167 & 0.6935 & 29  \\ \hline
20                        & 3369.0  & 0.5114  & 7   & 2957.1  & 0.4278  & 21  & 3194.7  & 0.6644  & 13  & 2603.880 & 0.6543 & 33  \\ \hline
40                        & 3276.0  & 0.5202  & 10  & 3021.0  & 0.4375  & 19  & 3549.1  & 0.8387  & 2   & 2666.044 & 0.6575 & 32  \\ \hline
80                        & 3369.8  & 0.5112  & 7   & 2951.5  & 0.4193  & 21  & 3596.0  & 0.9098  & 1   & 2740.896 & 0.6796 & 29  \\ \hline
\multicolumn{13}{|c|}{Conflict rate}                                                                                              \\ \hline
1                         & 3600.0  & 0.6718  & 0   & 3191.4  & 0.5689  & 13  & 3138.8  & 0.6773  & 15  & 2992.671 & 0.7465 & 21  \\ \hline
2                         & 3368.6  & 0.4972  & 7   & 2891.9  & 0.4269  & 23  & 3192.4  & 0.7253  & 13  & 2671.296 & 0.6581 & 30  \\ \hline
4                         & 2737.2  & 0.3579  & 27  & 2312.7  & 0.2880  & 42  & 3203.4  & 0.7601  & 13  & 1896.029 & 0.4837 & 53  \\ \hline
8                         & 3550.5  & 0.5254  & 2   & 3496.7  & 0.4344  & 4   & 3258.0  & 0.7376  & 11  & 3165.990 & 0.7967 & 19  \\ \hline
\multicolumn{13}{|c|}{Processing time settings}                                                                                   \\ \hline
1                         & 3362.4  & 0.5313  & 10  & 3047.6  & 0.4622  & 23  & 3583.0  & 0.8852  & 1   & 2891.071 & 0.7552 & 30  \\ \hline
2                         & 3328.7  & 0.5061  & 11  & 2933.7  & 0.4036  & 30  & 3594.8  & 0.8449  & 1   & 2558.680 & 0.6115 & 47  \\ \hline
3                         & 3251.2  & 0.5018  & 15  & 2938.2  & 0.4229  & 29  & 2416.6  & 0.4517  & 50  & 2594.739 & 0.6470 & 46  \\ \hline
\multicolumn{13}{|c|}{}                                                                                                           \\ \hline
Total                     & 3314.1  & 0.5131  & 36  & 2973.2  & 0.4296  & 82  & 3198.2  & 0.7252  & 52  & 2681.497 & 0.6712 & 123 \\ \hline
\end{tabular}
}
\caption{Comparison of different formulations for the $\mS$ problem}
\label{tab-p2}
\end{table}

Table \ref{tab-p2} reports the results for the $\mS$ problem. We derive several observations as follows:
\begin{itemize}
    \item In overall, CP still performs the best when it can solve to optimality 123 over 432 instances. The worst formulation is (F1). Formulation (F2) overcomes (F3) to become the second best formulation in terms of the number of successfully solved instances. Formulation (F3) can solve more instances to optimality than (CP) in two cases: the instances with 32 jobs and the ones with the third processing time setting.
    \item The performance of all the formulations tends to decrease when the number of jobs increases. Formulation (F3) can solve 11 instances of 32 jobs to optimality, while three formulations (F1), (F2), and (CP) fail to close the gaps for all these instances. 
    \item Unlike the first problem, the performance of (F1), (F2) and (CP) for the second problem is better when the number of machines increases. The impact of the number of machines on the performance of (F3) is not really clear.
    \item The machine capacity has less impact on the performance of formulations (F1), (F2), and (CP). But the high values of the capacity lead to significant reduction on the number of successfully solved instances for formulation (F3). This phenomenon is predictable because the number of variables and constraints of (F3) depends on the machine capacity. This observation is similar to the first problem.
    \item The performance of (F3) is insensitive with the conflict rate for (F3). For three remaining formulations, the instances with conflict rate $\alpha = 4$ seem to be the easiest.
    \item Setting 3 of the processing time tends to be easier to solve than the two other settings for formulations (F1) and (F3) whose number of variables and constraints rapidly increases when the value of $T_{max}$ is higher. The performance of (F2) and (CP) is insensitive to the processing time setting.
\end{itemize}

\subsection{Results for the $\MM$ problem}
We now compare the exact methods based on the four formulations: (F1-$\MM$), (F2-$\MM$), (F3-$\MM$), and (CP-$\MM$); and the two methods based on the binary search (F1-BS) and (F3-BS). The obtained results are shown in Table \ref{tab-p3} where Columns ``Iter" report the number of iterations the search is performed in the binary search algorithms. 
%\end{adjustbox}
%P3%
%\begin{adjustbox}{width=1.15\textheight, rotate =0, angle = 90, caption={Comparison of different formulations for problem (P1)}, float = table

\begin{table}[!]
\centering
\begin{adjustbox}{angle=90}
\begin{tabular}{|c|c|c|c|c|c|c|c|c|c|c|c|c|c|c|c|c|c|c|c|c|}
\hline
\multirow{2}{*}{Instance} & \multicolumn{3}{c|}{F1} & \multicolumn{3}{c|}{F2} & \multicolumn{3}{c|}{F3} & \multicolumn{3}{c|}{CP} & \multicolumn{4}{c|}{F1-BS}   & \multicolumn{4}{c|}{F3-BS}      \\ \cline{2-21} 
                          & Time    & Gap     & Opt & Time    & Gap     & Opt & Time    & Gap     & Opt & Time     & Gap    & Opt & Time   & Iter & Gap    & Opt & Time     & Iter  & Gap    & Opt \\ \hline
\multicolumn{21}{|c|}{Number of jobs}                                                                                                                                                              \\ \hline
16                        & 2333.9  & 0.2769  & 51  & 2146.8  & 0.2283  & 60  & 1204.4  & 0.1679  & 100 & 908.356  & 0.2025 & 108 & 969.6  & 3.1  & 0.1611 & 107 & 966.948  & 3.410 & 0.2281 & 109 \\ \hline
24                        & 2990.1  & 0.3262  & 31  & 2556.6  & 0.2833  & 45  & 2316.1  & 0.4294  & 56  & 701.881  & 0.0893 & 117 & 1885.7 & 2.8  & 0.0943 & 80  & 1995.529 & 1.958 & 0.8878 & 68  \\ \hline
32                        & 3556.8  & 0.4837  & 3   & 3198.4  & 0.4378  & 21  & 2694.0  & 0.5517  & 42  & 991.753  & 0.1502 & 105 & 2054.2 & 2.5  & 0.0788 & 67  & 2207.820 & 1.882 & 1.1087 & 60  \\ \hline
\multicolumn{21}{|c|}{Number of machines}                                                                                                                                                          \\ \hline
2                         & 3600.1  & 0.6038  & 0   & 3600.1  & 0.5636  & 0   & 2165.3  & 0.4015  & 65  & 507.402  & 0.1079 & 124 & 1137.5 & 2.6  & 0.0742 & 100 & 1631.355 & 2.313 & 0.1360 & 82  \\ \hline
4                         & 3115.8  & 0.3236  & 22  & 2620.9  & 0.2677  & 44  & 2101.5  & 0.3956  & 63  & 1080.833 & 0.1924 & 101 & 1944.2 & 2.6  & 0.1047 & 75  & 1808.256 & 2.313 & 0.7400 & 76  \\ \hline
6                         & 2164.8  & 0.1595  & 63  & 1680.9  & 0.1182  & 82  & 1947.7  & 0.3518  & 70  & 1013.755 & 0.1418 & 105 & 1827.9 & 3.3  & 0.1554 & 79  & 1730.686 & 2.625 & 1.3487 & 79  \\ \hline
\multicolumn{21}{|c|}{Machine capacity}                                                                                                                                                            \\ \hline
10                        & 2991.6  & 0.3950  & 20  & 2746.5  & 0.3454  & 29  & 882.7   & 0.0819  & 88  & 511.508  & 0.0963 & 93  & 1282.0 & 2.1  & 0.1128 & 76  & 455.089  & 2.019 & 0.1439 & 97  \\ \hline
20                        & 2914.5  & 0.3595  & 22  & 2593.7  & 0.3160  & 32  & 1860.7  & 0.3059  & 55  & 838.071  & 0.1496 & 83  & 1364.0 & 2.5  & 0.1129 & 72  & 1513.904 & 2.435 & 0.3693 & 66  \\ \hline
40                        & 2940.2  & 0.3519  & 22  & 2656.2  & 0.3078  & 32  & 2437.7  & 0.4914  & 37  & 954.645  & 0.1661 & 80  & 1859.1 & 3.1  & 0.1188 & 57  & 2195.810 & 2.546 & 0.9914 & 45  \\ \hline
80                        & 2994.7  & 0.3427  & 21  & 2539.3  & 0.2967  & 33  & 3104.9  & 0.6527  & 18  & 1165.095 & 0.1775 & 74  & 2041.0 & 3.6  & 0.1010 & 49  & 2728.927 & 2.667 & 1.4617 & 29  \\ \hline
\multicolumn{21}{|c|}{Conflict rate}                                                                                                                                                               \\ \hline
1                         & 3207.9  & 0.4434  & 12  & 3173.3  & 0.4233  & 13  & 1914.0  & 0.3359  & 55  & 634.180  & 0.0745 & 89  & 506.0  & 2.3  & 0.0033 & 96  & 1438.842 & 1.833 & 0.4285 & 67  \\ \hline
2                         & 2910.2  & 0.3540  & 21  & 2722.3  & 0.3293  & 28  & 1913.7  & 0.3673  & 54  & 622.403  & 0.0761 & 90  & 1023.2 & 2.7  & 0.0087 & 82  & 1501.743 & 2.074 & 0.5299 & 68  \\ \hline
4                         & 2442.2  & 0.2518  & 37  & 2119.8  & 0.2204  & 46  & 2324.3  & 0.4015  & 43  & 616.228  & 0.0949 & 90  & 1513.2 & 3.4  & 0.0210 & 67  & 1965.017 & 2.241 & 0.6854 & 52  \\ \hline
8                         & 3280.6  & 0.4000  & 15  & 2520.4  & 0.2929  & 39  & 2133.9  & 0.4272  & 46  & 1596.509 & 0.3440 & 61  & 3503.7 & 2.9  & 0.4126 & 9   & 1988.128 & 3.519 & 1.3225 & 50  \\ \hline
\multicolumn{21}{|c|}{Processing time   settings}                                                                                                                                                  \\ \hline
1                         & 3099.3  & 0.3894  & 21  & 2779.7  & 0.3483  & 36  & 2833.3  & 0.5840  & 34  & 1275.793 & 0.2153 & 94  & 2040.3 & 3.1  & 0.1367 & 69  & 2500.681 & 2.486 & 1.0661 & 47  \\ \hline
2                         & 2823.3  & 0.3202  & 35  & 2544.1  & 0.2792  & 45  & 2361.2  & 0.4472  & 55  & 847.179  & 0.1305 & 111 & 1625.5 & 3.3  & 0.0784 & 84  & 2053.826 & 2.604 & 0.9771 & 67  \\ \hline
3                         & 2958.1  & 0.3772  & 29  & 2578.0  & 0.3220  & 45  & 1019.9  & 0.1178  & 109 & 479.017  & 0.0964 & 125 & 1243.7 & 2.1  & 0.1192 & 101 & 615.790  & 2.160 & 0.1815 & 123 \\ \hline
\multicolumn{21}{|c|}{}                                                                                                                                                                            \\ \hline
Total                     & 2960.2  & 0.3623  & 85  & 2633.9  & 0.3165  & 126 & 2071.5  & 0.3830  & 198 & 867.330  & 0.1474 & 330 & 1636.5 & 2.8  & 0.1114 & 254 & 1723.433 & 2.417 & 0.7416 & 237 \\ \hline
\end{tabular}

\end{adjustbox}
\caption{Comparison of different methods for the $\MM$ problem}
\label{tab-p3}
\end{table}
From Table \ref{tab-p3}, several observations are derived as following:
\begin{itemize}
    \item The binary search algorithms significantly improve the original formulations. The CP formulation is still the best, followed by (F1-BS) and (F3-BS) in terms of the number of instances solved to optimality. However, we also observe that (F3-BS) is better then (CP) on instances with one of two characteristics: $n = 16$ or $T=10$ while (F1-BS) performs better than (CP) on instances with conflict rate set to 1. Among the three methods based on the pure MILP formulations, our new model (F3) is the dominant, followed by (F2) and (F1). Although formulation (F1) performs worse than formulation (F3), the F1-based binary search algorithm can solve more instances than the algorithm based on (F3).
    \item The number of jobs increases, the performance of MILP-based methods (including binary searches) decreases. The (CP) model is less dependent on the number of jobs.
    \item Four methods (F3), (CP), (F1-BS), and (F3-BS) can solve more instances with the least number of machines that is equal to 2. The performance of two formulations (F1) and (F2) increases when the number of machines increases. 
    \item The performance of (F1) and (F2) is insensitive with the machine capacity. The remaining methods can solve more instances to optimality when the machine capacity is smaller.
    \item The increase in the number of conflict jobs leads to harder instances for methods (CP), (F1-BS), and (F3-BS). The instances with more conflict jobs require more iterations to solve for the two binary search methods. The impact of the conflict rate on three pure MILP methods is less clear.  
    \item F3-based methods can solve more instances with the setting 3 of the processing time. The impact of the processing time settings on the performance of the other methods is not clear. 
\end{itemize}

\subsection{Other discussions}

We now discus the hardness of the three problems based on the performance of the algorithms. Surprisingly, the min-max objective function which has a non-linear characteristic does not make the $\MM$ problem the most challenging. Among three problems, the $\mS$ problem is the most difficult for the proposed methods, followed by the $\MM$ and $\MS$. 

In terms of the performance on the three problems, (CP) is in overall the best when it can solve more instances of all the problems than other methods. Among MILP-based instances, our new formulation (F3) performs better than two existing formulations (F1) and (F2) in terms of the number of solved instances.

 Although (CP) dominates other formulations, we also note that there are a number of instances that (CP) cannot solve to optimality while the MILP-based methods can. As can be seen in Table \ref{unSolveCP}, all the formulations can provide optimal solutions on several instances that (CP) cannot close the gaps. More precisely, (F1) and (F2) can solve 2 and 3 instances of the $\mS$ problem that are not solved to optimality by (CP). In particular, (F3) can perform better than (CP) on instances of all problems. Therefore, for the $\MS$ and $\mS$ problems, we recommend to use in practice the (CP) and (F3) models for exact methods. For the $\MM$ problem, the algorithms based on (CP), (F1-BS), and (F3-BS) could be selected options.

\begin{table}[h]
\centering
\begin{tabular}{|c|c|c|c|}
\hline
      & $\MS$ & $\mS$ & $\MM$ \\ \hline
F1    & 0  & 2  & 0  \\ \hline
F2    & 0  & 3  & 0  \\ \hline
F3    & 23 & 30 & 40 \\ \hline
F1-BS &    &    & 10 \\ \hline
F3-BS &    &    & 50 \\ \hline
\end{tabular}
\caption{The number of instances MILP-based methods can solve to optimality while CP cannot}
\label{unSolveCP}
\end{table}

\section{Conclusion}
\label{sec:conclude}
We study in this paper a class of three scheduling problems where some jobs cannot be scheduled concurrently on different machines: $\MS$, $\mS$, and $\MM$. We propose two new models based on mixed integer linear programming and constraint programming for the $\MS$ problem and adapt these models and two other existing ones for the $\mS$ and $\MM$ problems. Corresponding solvers are used to solve the models to optimality, leading the first exact methods proposed for the problems. In addition, binary search-based exact algorithms also are developed for the $\MM$ problem. Results of experiments carried out on instances with up to 32 jobs and 6 machines are reported and analysed. The performance of new models is clearly better than that of existing models in terms of the number of instances solved to optimality. The binary search algorithms for the $\MM$ problem also show their effectiveness.

The research perspectives are numerous. The MILP formulations could be strengthen with additional valid inequalities or optimality cuts to develop efficient branch-and-cut algorithms. Metaheuristics could also be developed to provide good solutions for larger instances in short computation time. Finally, other variants of scheduling problems (e.g., flow shop, job shop) with the conflict constraint are interesting topics for other researches.

\section*{Acknowledgment} This work has been supported by Phenikaa University.

\bibliographystyle{spmpsci}
\bibliography{reference}

\end{document}